\documentclass{icrc29}
\usepackage{graphicx,amssymb,amsmath,times}
\begin{document}
\title[Magnetospheric Eternally Collapsing Objects ...]
{Magnetospheric Eternally Collapsing Objects (MECOs): Likely New Class of Source
of Cosmic Particle Acceleration}
\author[Abhas Mitra] {Abhas Mitra 
        \newauthor\\
         Nuclear Research Lab., Bhabha Atomic Research Centre, 
            Mumbai 400 085, India\\        }
\presenter{Presenter: Abhas Mitra (amitra@apsara.barc.ernet.in), \  
ind-mitra-A-abs1-og12-oral}

\maketitle

\begin{abstract}
It is known that spinning pulsars could be source of VHE-UHE cosmic
particle acceleration. It is also conjectured that (fictitious) spinning
Black Holes (BH) could be sites of cosmic particle acceleration. However,
it has  been shown by Mitra\cite{1,2,3,4,5} and  Leiter and Robertson\cite{6} 
that General
Relativity (GR) actually {\em does not allow the existence or formation of finite
mass BHs}!  It was predicted that the BH Candiadates (BHCs) have strong intrinsic
magnetic fields (like pulsars) instead of Event Horizons. And this
prediction has tentatively been verified in a series of papers by
Robertson \& Leiter and Schild\cite{7,8,9,10}. Thus all observed BH Candidates are actually
not BHs, and, they are expected to be MECOs. In fact, ``MECO'' has been
listed as one of the frequently used abbreviations in current A\& A
literature ($www.phys.uni-sofia.bg/~astro.html$). Stellar mass MECOs are GR analogs of
conventionally known isolated pulsars. While pulsars are/have (i) COLD,
i.e., not supported by radiation pressure, MECOs are HOT, i.e., supported
primarily by trapped radiation pressure, (ii) upper mass limit of 3-4 $M_\odot$
 MECOs, being HOT, have no Upper Mass Limit, (iii) surface
gravitational red shift, $z\sim  0.1 -0.2$, MECOs have $z \gg 1$ so that photons
can remain almost permanently trapped inside them. It may be recalled
that isolated (non- accreting) uncharged BHs are cold and dead objects
without any physical activity. On the other hand, spinning MECOs are like
GR pulsars. 
\end{abstract}
\section{Introduction}
Cosmic Ray Acceleration may take place in several kinds of sites. 
One kind of sites could be of extended nature, e.g., Supernova
Remnants (SNRs), Pulsar Wind Schocks/Nebula (PWNs), or Jets.   The other kind
could direcly involve Compact Objects such as Spinning Neutron Stars, i.e., 
Pulsars and (fictitious) Spinning Black Holes. The accretion shocks around such compact objects
might also be the sites of acceleration even when such compact objects would not spin.
In particular, for extragalactic point sources, we invariably invoke massive BHs
as the Central Compact Object. But it is time now to have a fundamental revision
of this idea in the light of theoretical astrophysics/GR research carried out
in past few years. Although it will sound unthinkable, it is a hard fact that,
it has been unequivocally shown that the supposed BH candidates cannot be BHs
and they must be something else\cite{1,2,3,4,5,6}. Within the context of the classical GR,
this ``something else'' has turned out to  be Eternally Collapsing Objects (ECOs)
and their specific version: Magnetospheric Eternally Collapsing Objects (MECOs).
Even though the magnetic moment of MECOs (as measured by distant observers) 
could be same or even lower than typical young pulsar values, 
the spinning MECOs could beat the
pulsars as sources of acceleration of high energy ions/electrons by virtue of
extreming GR Frame Dragginf effect\cite{11}.
In the present paper, we would like to introduce ECOs/MECOs to cosmic ray astrophysicists
in general and also argue why ECOs are (rather than BHs) 
inevitable products of collapse of massive stars.
\section{Reasons for Non-existence of BHs}
It is true that the so-called vacuum Schwarzschild solution of GR {\em apparently}
makes a strong case of existence for BHs. This solution involves an {\em integration constant}
$\alpha_0$ which is interpreted as twice the mass of the BH:
\begin{equation}
\alpha_0 = {2 G M_0/c^2}
\end{equation}
where $G$ is the gravitational constant and $c$ is the speed of light. It has all along
been {\bf assumed} that this integration constant can have arbitrary finite value.
But by application of basic differential geometry (invariance of 4-volume),
 it has been shown recently, that,
actually, this integration constant is unique: $\alpha_0 \equiv 0$\cite{3}. This shows that,
the vacuum Schwarzschild solution is only  of notional value in the sense that
BHs have a unique mass $M_0 \equiv 0$, and as far as real objects ($M >0$) are
concerned, they cannot be BHs. What are the physical reasons for non-occurrence
of finite mass BHs?
\begin{enumerate}
\item
It turns out that the radius of the BH is none other than $\alpha_0$ and the
fictitious  boundary with $r=\alpha_0=r_g$ is called the ``Event Horizon'' (EH). In case there would
be a finite mass (and radius) BH, then the speed of any material particle
would become $v=c$ at the EH. Further one would have $v >c$ for $r < \alpha_0$.
This would be in violation of GR and hence there cannot be a region with $r < \alpha_0$.
In other words ,one must have $\alpha_0 = 2GM_0/c^2 \equiv 0$, implying $M_0\equiv 0$.
\item
The acceleration of any object as measured by any observer (i.e., acceleration scalar)
would be $a=\infty$ at the EH in tune with an infinite Lorentz factor at $r=\alpha_0$
And for $r < \alpha_0$, one would have $a=imaginary$(!!) just like the Lorentz
 factor would become imaginary if one would misconceive of a situation with $v >c$.
\item
It has all along been {\bf assumed} that if a massive star would collapse, somehow
a {\em trapped surface} would be formed - and then,
the formation of a BH would be inevitable.
But it has been shown, in a highly transparent and exact manner, that GR actually
does not allow formation of trapped surfaces\cite{4}.
\item
The physical reason for non-occurrence of trapped surfaces is again essentially the same:
Formation of a trapped surface corresponds to a speed of the collapsing fluid
$v \ge c$ in violation of GR.
\end{enumerate}
\section{Then What Happens?}
To appreciate this, one has to recall the definition of ``Gravitational Redshift''
$z$ associated with any spherical object of mass $M$ and radius $R$\cite{12}:
\begin{equation}
1+z = (1-2GM/Rc^2)^{-1/2}
\end{equation}
For the Neutron Stars, supposed to be highly compact, the value of $z$ lies
in the modest range of $0.1-0.2$. But for the EH of a Schwarzschild BH, one has
$z =\infty$. If one plots the  compact objects against $z$, there would be
and {\em infinite gap} between a NS and a BH! And it believed that during the
collapse, once the star crosses the limit of $z=0.1-0.3$ (NS), then it directly jumps to
$z=\infty$ (BH).
Obviously the star has to pass through the range of say $z=1$, $z=10$ and so on
before arriving at $z=\infty$. But it is known that, the escape angle $\psi$ of even photons/
neutrinos
(not to talk of matter) becomes extremely narrow for $z >2.0$ even in vacuum\cite{11}:
\begin{equation}
\sin \psi < {3\sqrt{3} G M\over r c^2} (1+z)^{-1}\approx 2.5 (1+z)^{-1};~~z \gg 1
\end{equation}
As a result, once the collapse is strong enough to overcome the NS stage, 
 for sufficiently massive stars, the radiation
generated during the collapse {\em becomes virtually trapped} within the collapsing body
as $\psi \to 0$ (almost).
It is quite likely that the luminosity of the 
trapped radiation attains the appropriate Eddington value at extremely high value of $z$.
And it is this {\em trapped radiation and the associated pressure which virtually halt the collapse}
to make a quasi steady ECO.

$\bullet$ Why Eternally Collapsing?
In a strict sense some radiation always escape at any finite $z$, and in the same sense, and an ECO
is always collapsing and losing mass-energy. It asymptotically tries to achieve the $z=\infty$
BH stage with $M \to M_0 \equiv 0$.

\subsection{What is a MECO?}
Robertson \& Leiter\cite{7,8,9} found that the X-ray as well as radio observations from all  stellar mass binary
 NSs and BHCs can
be understood in terms of the ECO model where the magnetic energy density is almost as strong
as gravitational energy density, $B \sim 10^{16-17}$ G, and ECOs with such super strong $B$ are called MECOs. For self-consistency, they argued that,
(observed) MECOs should have $z \sim 10^{7-8}$. If the reader would have difficulty in accepting such high values
of $z$, recall that BHs have $z=\infty$. The MECO model, and, in particular the idea that MECOs must be supported by
Eddington limited internal luminosity, is due to Robertson \& Leiter\cite{6,7,8}.

\section{ Some Expected Questions}
\begin{enumerate}
\item
What about the Exact Oppenheimer-Snyder Collapse Solution?

The exact solution of gravitational collapse is indeed due to Oppenheimer \&
Snyder. But they adopted a fluid (dust) whose pressure $p\equiv 0$. Thermodynamics
demands that $p\equiv 0$ is possible iff density $\rho\equiv 0$. Therefore, the
corresponding mass of the Oppenheimer-Snyder BH is $M =M_0 \equiv 0$ in accordance
with\cite{3}.
\item
What about the Upper Limit of Mass of Neutron Stars?

Both Chandrasekhar Limit and NS upper mass Limit correspond to {\bf cold} objects, i.e., objects
primarily supported by cold degeneracy pressure. Objects which are not primarily supported by
such degeneracy pressure may have any mass: e.g., there are individual stars with mass $\sim 100
M_\odot$, our galaxy has a mass $\sim 10^{11} M_\odot$. However such objects are not compact. Nonetheless,
{\em stable} supermassive stars, supported by radiation pressure, can have arbitrary high mass
and compactness somewhat smaller than NSs.

ECOs are GR version of  {\em unstable} radiation/magnetic pressure supported objects and 
GR very much admits
their existence by means of the {\em Vaidya Metric}\cite{6}. Because of intense internal radiation, the baryonic plasma of ECOs may not be
degenerate at all. 
\item
If BHs really are not there, how is it that almost all  GR experts and astrophysicists have been using it and getting
nice results too?

This question is essentially a sociological one and has been partly answered in the small
article : Why no ``New Einstein'' by Lee Smolin\cite{13}.

To be precise, not a single astrophysical reserch paper, to the knowledge of the present author,
 has so far used the exact 
EH bounday conditions, like $z=\infty$, $a=\infty$ in any exact astrophysical computation. Invariably,
they avoid the EH and work away from it, e.g., like the ``Membrane Paradigm''\cite{14} where one virtually
replaces the EH by a physical conducting surface above it! In fact, these authors 
admit (p.46)\cite{14}
that:

``The mental {\bf deceit} of stretching the horizon is made mathematically
 viable, indeed very attractive, by the elegant set of membrane-like 
boundary conditions to which it leads at the stretched horizon....''
(The emphasis is by the author)

If BHs were physical objects, there would not have been any 
 need for a ``mental deceit'' and they could have been
 treated exactly without the prop of a ``membrane''.
\end{enumerate}
\section{Conclusions}
The so-called BH candidates are most likely to be ECOs. An isolated spinning
ECO, by virtue of  intense GR frame dragging effect,
will generate an induced electric field which could be stronger by a factor of $\sim 10^4$ than that of conventional pulsars\cite{11}.
Although, Robertson \& Leiter have constructed reasonably self-consistent model
of ECOs (MECOs) with a range of $B \sim 10^{16-17}$G and $z\sim 10^{7-8}$ for observed stellar mass BHCs,
there could be ECOs with much lower $B$ and $z$ and which might be unidentified
at present. For various physical properties of MECOs/ECOs, the reader is referred
to\cite{10}.
It may be mentioned that as far as detail internal structure of a quasar is concerned, probably,
Q0591+561 makes the best case as it has been studied for 40 years. In particular, 25 years of
gravitational lensing and microlensing studies have revealed an internal structure which cannot be
understood by the BH paradigm. On the other hand, the MECO model has been successful in
fixing some of basic structure parameters\cite{10}.
More on ECO accelerators will be presented in an adjacent paper\cite{11}.
\section{Acknowledgements}
The author is thankful to Rudy Schild, Stan Robertson and Darryl Leiter for
various useful discussions on the MECO model.

\end{document}